\documentstyle[12pt]{article}
\renewcommand{\baselinestretch}{1.3}
\newcommand{\be}{\begin{equation}}
\newcommand{\ee}{\end{equation}}
\begin{document}

\date{}
\title{Semiempirical charge distribution of clusters in the ion
sputtering of metal}
\author{Victor I. Matveev and Olga V. Karpova\\
Heat Physics Department of Uzbek Academy of Sciences,\\
28 Katartal Str., 700135 Tashkent, Uzbekistan}

\begin{titlepage}
\maketitle

\begin{abstract}
We propose the generalization of a known established empirically
(Wahl W. and Wucher A. Nucl. Instrum. Meth. {\bf B 94}, 36(1994))
power law, describing relative mass-spectra of neutral sputtered clusters,
on the cases of arbitrary cluster charges.
The fluctuation mechanism of charge
state formation of sputtering products in the form of large
clusters with the number of atoms $N\geq 5$ is also proposed.
The simple formula obtained by us has been shown a  good
agreement with the experimental data.
\end{abstract}
\end{titlepage}

\large

Sputtering of solids under  the ion bombardment is one of the main
applied and fundamental problems which is importance in the many directions
of contemporary science and technology. Considerable technological
possibilities in the micro- and nanoelectronics, cosmic and thermonuclear
technologies have stimulated increase of the number of works devoted to
application and basic investigations of sputtering phenomenon (see, for
example, recent reviews [1-3] and references therein). The theoretical
description and estimations of sputtering processes are rather difficult
due to the multiparticle character of problem  both at the stage of ion
penetration to solid and at the stage of formation sputtering products which
consist of not only single target atoms but also polyiatomic particles, i.e.,
clusters. Presently, some perspectives on carrying out of "first principle"
calculations are connected (see, also, estimates [4,5]) with computer
simulation by molecular dynamics methods. However, such calculations
are complicated in technical plan, especially in the case of increasing
of the number of atoms in cluster and they are difficult for performing
excluding the authors of these calculations.. The maximal using of
possibilities of empirically established sputtering regularities is
reasonable in this case. For cluster sputtering so-called power  law for
relative neutral cluster  yield which was discovered experimentally
(see for instance [6]) could be most important. According to this the
norlmalized neutral cluster yield is described by the law $N^{\xi}$,
where $N$ is the number of atoms in the cluster and the parameter $\xi$
depends on bombardment conditions and target type.
One of most complex problems
is also process of charge state formation of
surface sputtering products. Considerable number of experimental and
theoretical works are devoted (see, for example, review [7]) to the
investigations of charge state formation of single atomic particles at the
surface scattering or sputtering of metal surface. On other
hand, the mechanism of charged structure formation of polyatomic particles
had been less investigated both theoretically and experimentally.
In this paper the generalization of well known empirically established
power law describing relative mass-spectra of neutral clusters for cluster
emission of arbitrary charges is offered. The fluctuation mechanism of charge
state formation of sputtering products in form of large clusters with the
number of atoms $N\geq 5$ is also proposed. Derived  simple formulas are in
a good  accordance with the experimental data.
  We use the old conception according to which large
clusters are emitted as a whole agglomerate in the form of block of atoms
(see also [8,9]) .
We will consider the probability $W_N$ of events corresponding to correlated
movement of  $N$ -atomic block as given. Let us determine the charged state
of the block of N-atoms. For this purpose we will follow the statistical
deriving of Saha-Lengmuir's formula [10], and assume that with moving off
of the cluster from the metal surface up to some distance (so-called
critical distance) the exchange between the electrons of metal conduction
zone and electrons of cluster atoms  is possible. When cluster moves away
from the metal surface to  the distance exceeding critical one, the electron
exchange stop unadiabatically. Further below saying about cluster electrons,
we will mean valence electrons only and corresponding aggregate of states we
will call the cluster conduction zone. We will also  assume that namely
between
the zones of metal and cluster the exchange is possible. Then average number
of electrons $\overline{n_{\tau}}$ on the energy electron level
$\varepsilon_{\tau}$ of cluster, according to the Fermi distribution, is
defined by
$
\overline{n_{\tau}}=\{\exp[(\varepsilon_{\tau}-\mu)/\Theta]+1\}^{-1}\;,
$
where $\Theta$ is temperature, $\mu$ is the chemical potential. Let us
denote via  $\overline{\Delta n_{\tau}^2}$ the average of square deviation
numbers of occupation $n_{\tau}$
from the equilibrium  $\overline{n_{\tau}}$ - values. Then
$\overline{\Delta n_{\tau}^2}=\overline{(n_{\tau}-\overline{n_{\tau}})^2}
=\overline{n_{\tau}}(1-\overline{n_{\tau}})$ [11]. Obviously, the average
number of electrons is
$\overline{N_e}=\sum_{\tau}\overline{n_{\tau}}\;$.
Let the number of electron in cluster conduction zone is  $N_e$. Then,
according to definition, the average of square deviation of number of
electron in cluster conduction zone from average value is
$
\overline{\Delta N_{e}^2}=\overline{(N_{e}-\overline{N_{e}})^2}
=\sum_{\tau}\overline{\Delta n_{\tau}^2}\;.
$
The cluster, having $N_e$ electrons in conduction zone, will be electrically
neutral, if $N_e=\overline{N_e}$, where $\overline{N_e}$ is the average
number of electrons in the cluster conduction zone which is equal to the
number of atoms in $N$ -atomic cluster multiplied to valency $\gamma$
(i.e., to the  number
of atomic electrons, yelding  by neutral metal atom to the conduction zone).
Thus, cluster charge is  $Qe=(N_e-N\gamma)e$, where $e$ is electron charge.
Further calculations with these formulae require knowledge of the electronic
structure of cluster and generally speaking cannot be performed in general
form. However, if to consider cluster size is large enough and electronic
states are quasi-continuous, one can exchange summing over the electronic
states on integration over the zone [11]. Therefore, for the temperatures
less than the degeneration temperature, i.e. for
$\mu/\Theta \gg 1$, one has
$$
\overline{\Delta N_{e}^2}\approx
2^{1/2}\frac{Vm_e^{3/2}} {\pi^2\hbar^3}\sqrt{\mu}\Theta\;,
$$
where $m_e$ is electron mass in conduction zone,
$V$ is cluster volume
 and chemical potential of the degenerated Fermi gas with the number of
 particles
$\overline{N_e}$ in the cluster volume
 $V$ is [11]
$$
\mu=(3\pi^2)^{\frac{2}{3}}\frac{\hbar^2}{2m_e}
\left(\frac{\overline{N}_e}{V}\right)^{\frac{2}{3}}.
$$
Thus, the average of square deviation of cluster charge from the
equilibrium value of $\overline{Qe}=\overline{(N_e-N\gamma)e}=0$,
is
\footnote{In principle, equality to zero of equilibrium cluster charge
follows from the assumption that Fermi levels in cluster and metal coincide.
If it is not executed, asymmetry between positive and negative charged
clusters will be observed and corresponding changes in following formulas
can be easy made.}
\be
\overline{(\Delta Q_N)^2}=e^2\overline{\Delta N_{e}^2}=
e^2\frac{ 3^{\frac{1}{2}}}{\pi^{\frac{4}{3}}}
\frac{m_e \Theta}{\hbar^2}
\left(\frac{\overline{N}_e}{V}\right)^{\frac{1}{3}}V.
\label{eq:f12c}
\ee
 Probabilities $P_N(Q)$ of values $Q$
we will determine by making use of standard formula for probability of
fluctuations, i.e.,
\be
P_N(Q) =\frac{1}{D_N}
exp \biggl\{-\frac{1}{2}\frac{Q^2}{\overline{(\Delta Q_N)^2}} \biggl\},
\label{eq:f11c}
\ee
where normalizing factor  $D_N$ is defined by summing (2) over all possible
values (~\ref{eq:f11c}) $Q=0,\pm e,\pm 2e,...$.
Thus, to obtain probability $W_{N}^Q$ of cluster emission with
number of  atoms $N$ and charge
$Qe$ one should multiply the probability of occurrence of events $ W_{N} $
corresponding to correlated moving of $N$ -atomic agglomerate, on $P_N(Q)$:
\be
W_{N}^Q =  W_{N} P_N(Q).
\label{eq:f15c1}
\ee
On other hand, according to experiment, neutral clusters are distributed by
power law $N^{\xi}$,  and so
\be
W_{N}^{(Q=0)} =  W_{N} P_N(Q=0)=N^{\xi}.
\label{eq:f15w1}
\ee
Thus $W_{N}^Q$ can be written as follow

\be
W_{N}^Q =  \frac{1}{P_N(Q=0)}N^{\xi}P_N(Q).
\label{eq:f15w2}
\ee
As  $P_N(Q=0)=1/D_N$, then definitive expression for probability of
$N$-atomic cluster
emission  and having charge  $Q$ will have a form
\be
W_{N}^Q = N^{\xi}
exp \biggl\{-\frac{1}{2}\frac{Q^2}{\overline{(\Delta Q_N)^2}} \biggl\},
\label{eq:f19}
\ee
where, according to  equation (~\ref{eq:f12c}),
\be
\overline{(\Delta Q_N)^2}
= e^2\frac{ 3^{\frac{1}{2}}}{\pi^{\frac{4}{3}}}\frac{m_e \Theta}{\hbar^2}
\left(\frac{1}{d}\right)^{\frac{2}{3}}\gamma^{\frac{1}{3}}N\;,
\label{eq:f123}
\ee
where  $d$ is the number of atoms in the  unit of cluster volume, i.e.
concentration (which we have accepted equal to the  atomic target
concentration for numerical calculations).

Simplest characteristic of cluster charge distribution, consisting of given
number of atoms $N$,  is the ionization coefficient $\kappa^Q_N$
which is equal to the ratio of number of clusters with charge $Q\not=0$
and  number of neutral clusters with the  same number of  atoms $N$.
In our case  ionization coefficient is
\be
\kappa_N^{Q}=\frac{W_{N}^Q}{W_{N}^{Q=0}}=
exp \biggl\{-\frac{1}{2}\frac{Q^2}{\overline{(\Delta Q_N)^2}} \biggl\}.
\label{eq:f20}
\ee
Obviously, our consideration  is not applicable for the sputtering of single
atoms or small clusters. From comparison with the experimental data one can
made a conclusion (see also [8,9]) on applicability  of the model beginning
from the
concrete number of cluster atoms ( $N \geq 5$).
In experiment one measures, usually,  the relative probabilities of the
cluster yield with different number of atoms.
Therefore, to compare theoretical data with the experiment ones, one should
at first divide  the probability (6) to the probability of cluster emission with
 (~\ref{eq:f19}) $N=5$ (more exactly, we can choose any value  $N\geq5$, but it is
more conveniently for us, when
$N=5$ ) , i.e.
$Y^Q_N=W_{N}^Q/W_{5}^Q$.  The experimental data will be same
normalized. Farther, if it is necessary, one can pass to  arbitrary
convenient system of units.
        The results of analysis of the general formulas and performed
numerical calculations and experimental data which are given in
Figs. 1-3 allow to come to the following conclusions:
a) The charge state changes by the variation of target temperature,
moreover the ionization coefficients increase by increasing of the
temperature;
b) relative mass-spectra of the neutral clusters do not depend on target
temperature, while relative mass-spectra of charged clusters depend on it
very strongly, but by increasing of temperature they approach to
mass-spectra of neutral clusters;
c) the more cluster charge, the more seldom they are found; for example,
the number of clusters with charge 2, as a rule, less than the number of
clusters with charge 1;
d) large clusters are ionized in larger degree;
e) tendency to saturation of ionization coefficients with growth of cluster
dimension is an important peculiarity, qualitatively the same behavior has
been noted in experiments [12],  that confirms the conclusions about
coincidence of the relative mass-spectra of charged clusters with neutral
ones, when values of N are large (i.e., when $N \gg 1$) ).
As it is well known, the experimental registration of the charged clusters
is simpler technically than  one of neutral clusters. Therefore the data of
measurements of charged clusters allow restoring of neutral clusters
distribution indirectly and experimental set up is simplified very much.
\newpage
{\bf References } \\
1. H.H. Andersen, K.Dan. Vidensk. Selsk. Mat. Fys. Medd. {\bf 43}, 127(1993).\\
2. H.M. Urbassek and W.O. Hofer, K.Dan. Vidensk. Selsk. Mat.
Fys. Medd.       {\bf 43},  97(1993).\\
3. G. Betz and W. Wahl,
International J. of Spectrometry and Ion Processes. {\bf 140}, 1(1994).\\
4. A. Wucher and B.Y. Garrison, J.Chem Phys.
      {\bf 105}, 5999(1996). \\
5. Th.J. Colla, H.M. Urbassek, A. Wucher, C. Staudt,
R. Heinrich, B.J. Garrison, C. Dandachi and G. Betz
Nucl. Instrum. Meth. (1998), {\bf B 143}, 284(1998).\\
6. A. Wucher and W. Wahl, Nucl. Instrum. Meth.
       {\bf B 115}, 581(1996).\\
7. M.L. Yu, Topics of Applied Phys. Sputtering by Particle
Bombardment III. Ed. by R. Behrisch and K. Wittmaack, Springer-Verlag,
(1991) p. 91-160.\\
8. V.I. Matveev, S.F. Belykh and I.V. Veryovkin,
   Zh. Tekh. Fiz. {\bf 69}, 64(1999), [Technical Physics,
   {\bf 44}, 323(1999).].\\
9. S.F. Belykh, V.I. Matveev, I.V. Veryovkin, A. Adriaens, F. Adams.
   Nucl. Instrum.  Meth. {\bf B 155}, 409(1999).\\
10. Dobretsov L.N., Gomounov M.V. Emissional electronics,
Moscow, Nauka, 1966.\\
11. L.D. Landau and E.M. Lifshitz, Statistical physics, Part 1,
Moscow, Nauka, 1964.\\
12. W. Wahl and A. Wucher,
Nucl. Instrum. Meth. (1994), {\bf B 94},  36(1994).\\
13. S.F. Belykh, U.Kh. Rasulev, A.V. Samartsev and I.V. Veryovkin,
Nucl. Instrum. Meth. {\bf B 136-138},  773(1998).
\newpage
{\bf Figure captions:}\\

 Fig.1. The dependence of coefficients of single and double ionization
of clusters from 5 and 10 Ta-atoms on target temperature $\Theta$. \\

 Fig.2. The dependence of coefficients of single ionization on the number
of atoms in cluster of $Ag$: dotted line - our calculations at target
temperature
$\Theta=500^oK$,
$\bullet$ - experimental data from [12].\\

Fig.3. Relative yield $Y_N^1$ of one charge cluster of $Ta_N^{+1}$ in
dependence on number $N$ of atoms in cluster under one-charged ion of
$Au^{-1}$ (with the energy 6 keV) bombardment of tantalum at target                      and target
temperature $\Theta=2273^oK$:
unbroken line - calculated values of $Y_N^1$,
$\bullet$ - experiment [13].\\

\end{document}